\documentclass[conference]{IEEEtran}
\IEEEoverridecommandlockouts
\usepackage{cite}
\usepackage{amsmath,amssymb,amsfonts}
\usepackage[normalem]{ulem}
\usepackage{algorithmic}
\usepackage{graphicx}
\usepackage{textcomp}
\usepackage{xcolor}
\usepackage{float}
\def\BibTeX{{\rm B\kern-.05em{\sc i\kern-.025em b}\kern-.08em
    T\kern-.1667em\lower.7ex\hbox{E}\kern-.125emX}}
    
\usepackage{subcaption}
\usepackage{hyperref}

\newcommand{\fref}[1]{Fig.~\ref{#1}}

\newcommand{\tref}[1]{Table~\ref{#1}}

\IEEEoverridecommandlockouts
\IEEEpubid{\makebox[\columnwidth]{978-1-7281-9975-7/20/\$31.00~\copyright2020 IEEE \hfill}
\hspace{\columnsep}\makebox[\columnwidth]{}}

\begin{document}

\title{A Retinex based GAN Pipeline to Utilize Paired and Unpaired Datasets for Enhancing Low Light Images}

\author{
\IEEEauthorblockN{
Harshana Weligampola\IEEEauthorrefmark{1}, Gihan Jayatilaka\IEEEauthorrefmark{1},
Suren Sritharan\IEEEauthorrefmark{1}, \\
Roshan Godaliyadda\IEEEauthorrefmark{2}, 
Parakrama Ekanayaka\IEEEauthorrefmark{2},
Roshan Ragel\IEEEauthorrefmark{1},
 Vijitha Herath\IEEEauthorrefmark{2}}
\IEEEauthorblockA{ \{\IEEEauthorrefmark{1}\textit{Dept. of Computer Engineering}, \IEEEauthorrefmark{2}\textit{Dept. of Electrical and Electronics Engineering}\}\textit{, University of Peradeniya}\\\textit{Peradeniya 20400, Sri Lanka} \\
\{harshana.w, gihanjayatilka, suren.sri, roshangodd, mpb.ekanayake, roshanr, vijitha\}@eng.pdn.ac.lk
}
}
\newcommand\blfootnote[1]{%
  \begingroup
  \renewcommand\thefootnote{}\footnote{#1}%
  \addtocounter{footnote}{-1}%
  \endgroup
}

\maketitle

\begin{abstract}
\blfootnote{The published version of this paper is available at  \href{https://doi.org/10.1109/MERCon50084.2020.9185373}{https://doi.org/10.1109/MERCon50084.2020.9185373}}
Low light image enhancement is an important challenge for the development of robust computer vision algorithms. The machine learning approaches to this have been either unsupervised, supervised based on paired dataset or supervised based on unpaired dataset. This paper presents a novel deep learning pipeline that can learn from both paired and unpaired datasets. Convolution Neural Networks (CNNs) that are optimized to minimize standard loss, and Generative Adversarial Networks (GANs) that are optimized to minimize the adversarial loss are used to achieve different steps of the low light image enhancement process. Cycle consistency loss and a patched discriminator are utilized to further improve the performance. The paper also analyses the functionality and the performance of different components, hidden layers, and the entire pipeline.
\end{abstract}


\section{Introduction}
Recently there has been a great and growing interest in the field of computer vision and image processing. Due to the abundance of visual data and the higher computational power of portable devices, computer vision applications have been integrated into our day-to-day lives \cite{computer-vision-survey}.
However, many of the researches done on image transformation and interpretation are focused on well-lit images. Well-lit images can only be obtained under natural lighting during a fraction of the day or in the presence of artificial lighting during the rest of the day. There are situations where artificial lighting is not feasible (energy concerns, environmental concerns, obstruct natural lightning, etc.). Thus it is essential to identify low light images and enhance their features to guarantee the robustness of computer vision algorithms.

Low light images suffer from a range of issues such as low visibility, noise, colour distortion, etc. These issues prohibit them from being visually perceptible for the human eye and being useful for computer vision algorithms in terms of information richness. Enhancement of these low light images has been tried on hardware front (sensitive light-capturing mechanisms, longer exposure times, etc.) with varying degrees of success but at higher costs. The algorithmic approach to enhancing these images is considered to be an important research problem.

Algorithmic approaches to low light image enhancement are built upon the understanding of photonics, digital image properties, and biology related to human vision. Retinex theory of human vision \cite{Land1977} has been the defacto inspiration for most of the algorithmic pipelines in this domain. The algorithms can be broadly broken down into non-trainable algorithms which have a fixed activity based on programmer's understanding of the process, trainable (learning) algorithms which learn the low light image enhancement from example images, and hybrids of both these approaches (which are most successful). When it comes to learning algorithms (or pipelines utilizing trainable sections) we see systems built with trainable filters on standard CNNs \cite{recent-advances-cnn} and GANs \cite{gan}. 

These learning algorithms learn from datasets of two forms -- paired (containing a low light and well-lit captures of the same scene/object) and unpaired (containing sets of low light and high light images without a counterpart). The paired datasets are more informative but are difficult to obtain. They are relatively rare in number and the variety of scenes in them. In contrast, unpaired datasets are not as informative as the low light datasets, are easier to obtain, abundant, and contain a wide variety of scenes/objects. Existing solutions depend on only one of these two types.

This paper presents a DNN based algorithmic pipeline to enhance low light images with the following advantages.
\begin{itemize}
    \item The system utilizing both paired and unpaired datasets. Standard CNNs are used to learn from the paired datasets while GANs are used to learn from unpaired datasets.
    \item The CNN and GAN architectures and their coordination are designed with the retinex model as an inspiration.
\end{itemize}
\IEEEpubidadjcol
\section{Related works}
\subsubsection*{\textbf{Classical Algorithms}}
Classical image processing algorithms are unsupervised algorithms that enhance low light images through well-formulated mathematical models.
These algorithms are based on two schools of thought, namely intensity-based enhancement\cite{AHE} and gradient-based enhancement\cite{grad_enhance}. They are computationally efficient and simple.
But they are not robust enough to be used over different conditions without manual calibration.

\subsubsection*{\textbf{Retinex theory}}
The Retinex Theory \cite{Land1977}
is a biologically-motivated theory based on the colour constancy property of the human visual system (HVS). It states that any object has a lighting independent property known as the reflectance and the perceived colour of the object occurs due to the illumination on these objects. Thus, the retinex based algorithms focus on the decomposition of the image to obtain the reflectance which represents the ``true colour'' of the object.
Many variations of the original retinex model have been proposed.\cite{Land1977, land1983recent, wang2015variational} 
Recent works have made improvements to the retinex model for better performance and improved robustness \cite{ret_noise} while other techniques such as LIME \cite{lime} employs a similar idea for the decomposition and enhancement of illumination maps. The model proposed in this work is also motivated by the retinex model.


\subsubsection*{\textbf{Deep Convolution Networks}}
Deep learning techniques such as convolution neural networks (CNNs) \cite{lecun1999object} and autoencoder networks \cite{Ballard1987ModularLI} have been proved to perform better than classical algorithms in many image processing tasks \cite{imagenet,cnn1}. In the field of low-light image enhancement, neural networks have been proposed in works such as LLNet \cite{llnet} (based on stacked auto-encoder), LLCNN \cite{llcnn} (based on the retinex theory), MSRnet \cite{msrnet} (based on log transforms) and ~\cite{pipeline-nn} (based on wavelet transform). Recently, the Retinex-net \cite{deep-retinex-decomp} and \cite{see_in_dark} have been proposed. These are full frameworks for low light image enhancement, which brings impressive results. However, these techniques have certain shortcomings, the major one being that these architectures depend on training data in the form of paired datasets, which is difficult to obtain.

\subsubsection*{\textbf{Adversarial Networks}}
Generative Adversarial Network \cite{gan} have proven to perform sufficiently well for many supervised and unsupervised learning problems.
In \cite{cycle-gan} the authors propose a model through which the need for paired images has been elevated and image translation between two domains can be done through cycle-consistence loss. These techniques have been applied to many other applications including dehazing, super-resolution, etc. Lately, it has been applied to low light image enhancement in EnlightenGAN \cite{en-gan} with promising results and this has motivated our GAN model.


\subsubsection*{\textbf{Dataset}}
The main requirement for many learning models is the existence of paired training data with low/normal-light images and several methods exist to collect such data. The images in LOL \cite{deep-retinex-decomp} dataset are captured in the daytime under normal light condition by changing the exposure time and ISO.
The SID \cite{see_in_dark} dataset is generated by increasing the exposure time to generate well-lit images. However, it consists of raw sensor data under extremely low-light scenes, and this may limit its application for general low-light enhancement researches. Furthermore, due to the tedious experimental procedure, these datasets contain only a few images.



Supervised learning models depend on the availability of a large volume of training data but, the aforementioned techniques are neither efficient nor scalable. Thus synthetic image datasets are used as an alternative in many works \cite{brightening-train, attention_guide}. However, synthetic low light images are created by processing images taken under normal light conditions, and these algorithms may not consider other factors. This results in performance variation on low-light images taken in real-world conditions and therefore may not be suitable for non-synthetic dark images.

The introduction of learning models such as cyclic-GANs has removed the dependency on paired datasets.
Therefore, low light images present in \cite{Exdark, deep-retinex-decomp} \footnote{LOL contains paired images, but images can be sampled without the need to keep any pair} can be mixed with high exposure images to create a dataset of unpaired low/normal light image.




\section{Proposed Method}

The proposed model splits the enhancement processes into multiple stages and each stage should fulfil one of the objectives specified below.
\begin{itemize}
    \item The image's light level should be measured.
    \item The image's information should be extracted even in the poorly-light condition.
    \item The image's light should be increased while preserving and enhancing the information.
    \item The noise and deformations introduced to the image during the enhancement process should be taken care of.
\end{itemize}
This work proposes a pipeline to achieve all those objectives through a series of neural networks. The first two objectives are achieved through the \textbf{decomp-net} (motivated by the retinex theory) which decomposes the images into the reflectance (colour information) and illumination (lighting information). The next two objectives are achieved through the \textbf{enhancement-net}. The first part of the process focuses on local neighbourhood of pixels while the second part focuses on both the local and global neighbourhoods.

\begin{figure*}
    \centering
    \includegraphics[width=0.9\linewidth]{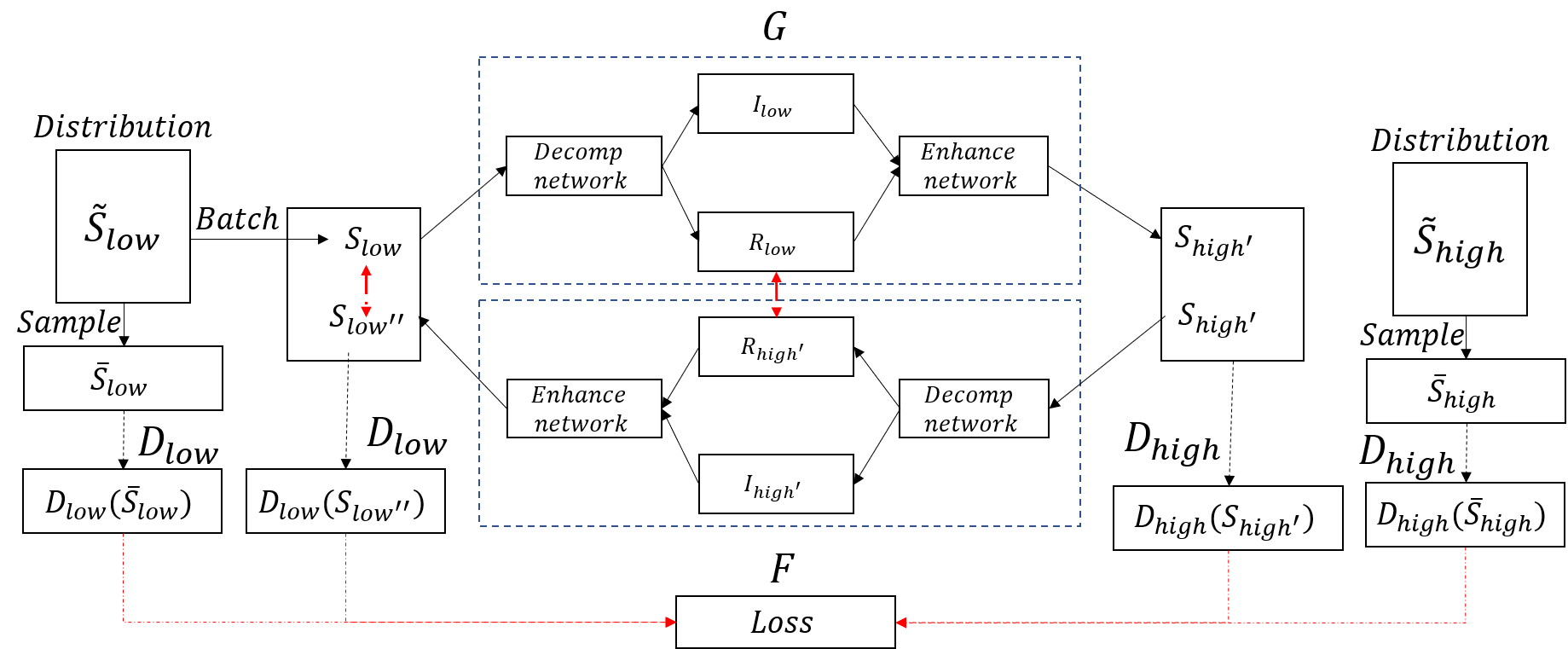}
    \caption{Forward cycle of the retinex-cycle-GAN model}
    \label{fig:full-system}
\end{figure*}

The decomposition requires an understanding of how the images change under different lighting conditions. This relationship could be learned effectively using paired images (same scene captured under well-lit and low light condition). The enhancement requires an understanding of different types of scenes and objects in images. It is difficult to obtain a paired dataset for this task. Therefore we depend on an unpaired dataset.
The overall architecture is given in \fref{fig:full-system} and the following subsections will describe the individual components.

\subsection{Retinex decomposition network}
The first part of the pipeline performs retinex decomposition on the images.
This retinex idea could be extended to digital images (where $S$ and $R$ are 2 spatial dimensional matrices with 3 colour channels and $I$ is a 2 spatial dimensional matrix with 1 channel) as $S = R \circ I$ where $\circ$ is the spatial element-wise multiplication. 
The images are decomposed into two parts as the reflectance $R$ and the illumination $I$.

\begin{itemize}
    \item Reflectance $R$ : This part has the colour information of the image. This is a 3 channel image of dimensions similar to the original image. We assume this part to be consistent over different lighting conditions for a single scene/object.
    \item Illumination $I$: This is a single channel image that represents the lighting effect of an image. 
\end{itemize}

We define the generators used in the translation  $S_{low} \rightarrow S_{high}$ as $G_i$ and the generators used in the translation $S_{high} \rightarrow S_{low}$ as $F_i$. We define the NN decomposition of low light images as $G_1$ and the NN decomposition of high light images as $F_1$. But the symmetric training process ensures that $G_1 = F_1$. The architecture is shown in \fref{fig:full-system} and \fref{fig:forward_reverse}. 

\begin{itemize}
    \item $G_1 : S_{low} \rightarrow [R_{low},I_{low}]$
    \item $F_1 : S_{high} \rightarrow [R_{high},I_{high}]$
\end{itemize}
For this, we use Deep Retinex Decomposition Network proposed in \cite{deep-retinex-decomp}. Both $G_1$ and $F_1$ were trained using LOL dataset \cite{deep-retinex-decomp} which consists of well-lit and low light, coloured image pairs. 16 random patches of size $64\times64$ are sampled from an image. The NN was trained to minimize the custom loss defined in \cite{deep-retinex-decomp} with the adam\cite{adam} optimizer at a learning rate of 0.001 and a decay factor of 0.9 for 100 iterations.

\subsection{Enhancement network}
The enhancement is done using a neural network. The architecture is inspired by U-Net \cite{unet} and its modifications over U-Net are given in \tref{tab:unet}. This CNN operates on the concatenation of $R$ and $I$ as the input (unlike U-Net). This will output an enhanced version of the illumination map. Since it is difficult to find a dataset with low light and well-lit image pairs, using a paired dataset learning technique to train the enhancement network is not practical. Therefore, we use an unpaired dataset with a Generative Adversarial Network (GAN) to train the enhancement network.
\begin{table}[h]
    \centering
    \caption{U-Net Specifications}
    \begin{tabular}{|c|c|}
    \hline
        Parameter & Value\\
        \hline
        Input size & $256 \times 256$ \\
        Down/upsampling factor & $\frac{1}{2} , 2$\\
        No. down/upsampling layers & $7 , 7$\\
        No channels (hidden layers) & 4,128,256,512,512,512,512,512\\
        Conv kernel size & $3 \times 3$\\
        \hline
    \end{tabular}
    \label{tab:unet}
\end{table}



We use the cycle consistency loss \cite{cycle-gan} to train the enhancement network with an unpaired dataset.

\subsubsection{Cycle consistency}
To enhance the training process we use two GANs\cite{gan} that will generate well-lit images from low light images and vice versa. We use these GANs as illustrated in \fref{fig:full-system} to preserve the cycle consistency using an adversarial loss \cite{cycle-gan}. This reduces the distance between the generated image and the expected image distributions.

The ``cycle'' consists of a forward pass $S_{low} \rightarrow S_{high'} \rightarrow S_{low''}$ and a backward pass $S_{low} \rightarrow S_{high'} \rightarrow S_{low''}$. This is further explained in \fref{fig:forward_reverse}, \fref{fig:full-system} and the list of equations given below.
\begin{itemize}
    \item $G_2 : [R_{low},I_{low}] \rightarrow I_{high'}$
    \item $G_3 : [R_{low},I_{high'}] \rightarrow S_{high'}$
    \item $S_{high'} = G(S_{low}) \approx S_{high}$
    \item $F_2 : [R_{high'},I_{high'}] \rightarrow I_{low''}$
    \item $F_3 : [R_{high'},I_{low''}] \rightarrow S_{low''}$
    \item $S_{low''} = F(S_{high'}) = F(G(S_{low})) \approx S_{low}$
\end{itemize}

$G_1, G_2, F_1$ and $F_2$ are trainable components (based on U-Net architecture) while $G_3$ and $F_3$ are non-trainable components.
\begin{figure}[htb!]
    \centering
    \begin{subfigure}[t]{\linewidth}
        \centering
    	\includegraphics[width=.9\linewidth]{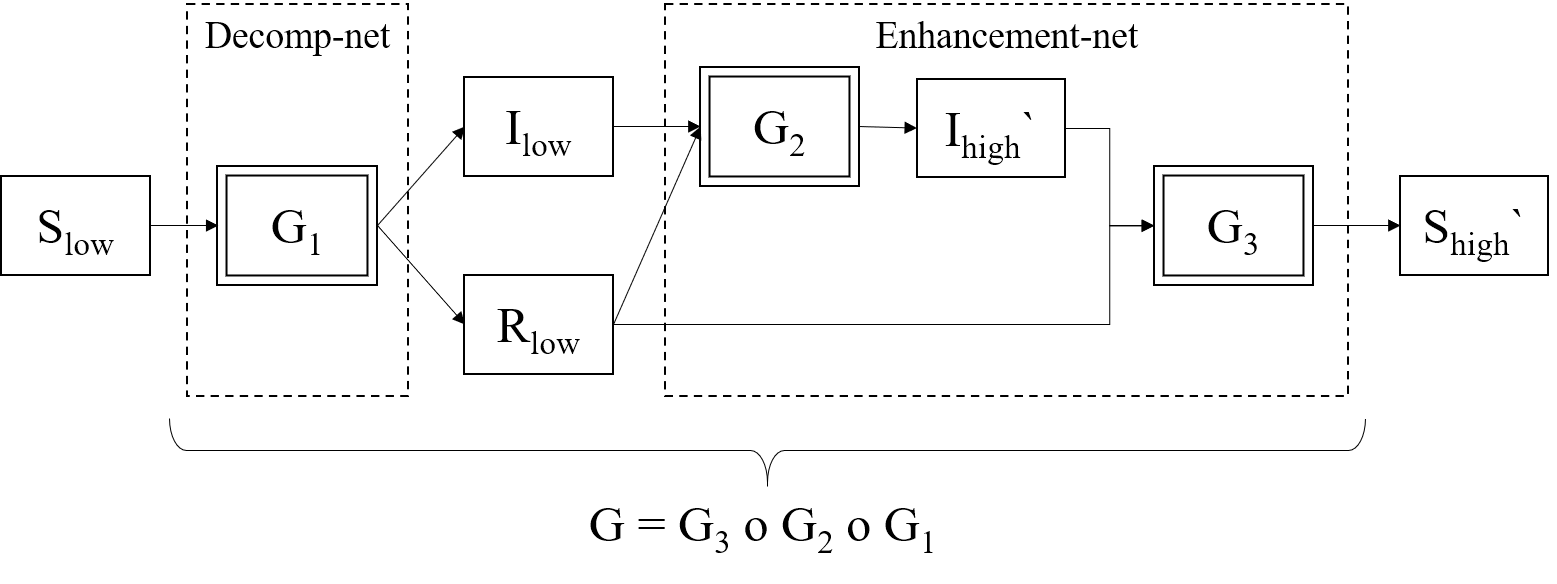}
        \caption{Forward generator.}
		\label{fig:forward}
    \end{subfigure}
    \begin{subfigure}[t]{\linewidth}
        \centering
    	\includegraphics[width=.9\linewidth]{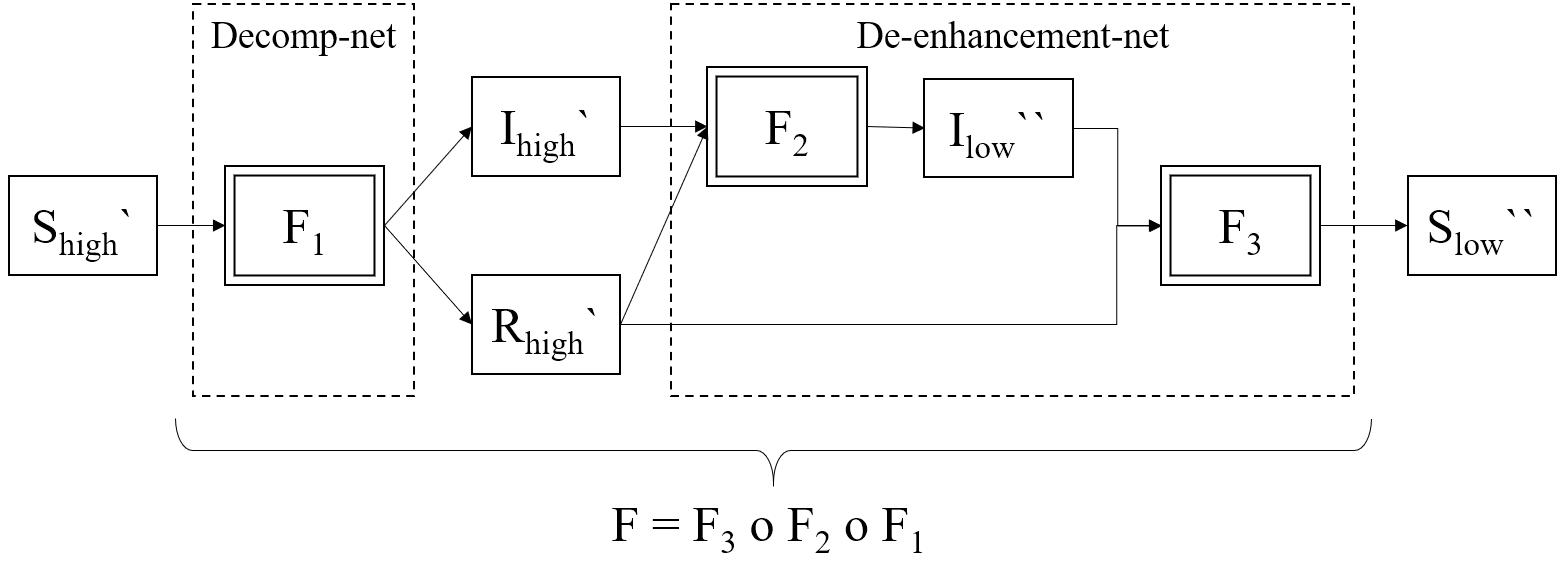}
        \caption{Forward generator.}
    	\label{fig:reverse}
    \end{subfigure}
    \caption{Block diagram of the generator components.}
    \label{fig:forward_reverse}
\end{figure}

To preserve the cycle consistency of the forward cycle $S_{low}$ and $S_{low''}$ should be the same. When considering the backward cycle, $S_{high}$ and $S_{high''}$ should be the same.
Thus, we include a cycle consistency loss by taking the difference between $S_{low}$, $S_{low''}$ and $S_{high}$, $S_{high''}$. as shown in \eqref{eq:cycle_loss_s}. In addition, using the retinex theory we know that reflectance map in each forward/backward cycle should be equal as well \cite{deep-retinex-decomp}. Therefore, we include the difference between $R_{low}$, $R_{high'}$ and $R_{high}$, $R_{low'}$ in the cycle consistency loss as well.  This loss is given in \eqref{eq:cycle_loss_r}. Then total cycle consistency loss is given by \eqref{eq:cycle_loss}

\begin{equation}
    \label{eq:cycle_loss_s}
    \centering
    \begin{aligned}
    \mathcal{L}_{cyc_S} = 
    &\,\mathbb{E}_{S_{low}~p(S_{low})} \left[ || F(G(S_{low})) - S_{low} ||_1 \right] +\\
    &\,\mathbb{E}_{S_{high}~p(S_{high})}[|| G(F(S_{high})) - S_{high} ||_1]
    \end{aligned}
\end{equation}

\begin{equation}
    \label{eq:cycle_loss_r}
    \mathcal{L}_{cyc_R} = || R_{low} - R_{high'} ||_2 + || R_{high} - R_{low'} ||_2
\end{equation}
\begin{equation}
    \label{eq:cycle_loss}
    \mathcal{L}_{cyc} = \mathcal{L}_{cyc_S} +\mathcal{L}_{cyc_R}
\end{equation}

\subsubsection{Training of enhancement network}

The total loss for the enhancement networks G and F including the generator loss is given in \eqref{eq:total_g_loss} and \eqref{eq:total_f_loss} where $H(p,q)$ is the binary cross entropy of distribution $q$ relative to a distribution $p$.
\begin{equation}
    \label{eq:total_g_loss}
    \mathcal{L}_G = \mathcal{L}_{cyc} + H(D_{high}(G(S_{low})), \mathbf{1})
\end{equation}
\begin{equation}
    \label{eq:total_f_loss}
    \mathcal{L}_F = \mathcal{L}_{cyc} +  H(D_{low}(F(S_{high})), \mathbf{1})  
\end{equation}

These NNs ($G_2, G_3, F_2, F_3$) were trained against a dataset created by adding 485 well-lit images from LOL dataset. The generators were trained to minimize the loss \eqref{eq:total_g_loss} and \eqref{eq:total_f_loss}. We use the Adam optimizer\cite{adam} at a learning rate of 0.0002 for the discriminator and generator. A decay factor of 0.5 is used. Each model is trained for 500 epochs with batch size 8.

\subsection{Patched Discriminator}

Usually, GANs use a discriminator to distinguish between real and generated data. When generating images, the whole image is processed and one scalar value between 0 and 1 is predicted. This method is not feasible for this problem because some regions in the image are more specific than the others. For example, an image with low-light in a small area can be identified as a well-light image by the discriminator if the whole image is considered.

\begin{figure}[tb]
    \centering
    \includegraphics[width=\linewidth]{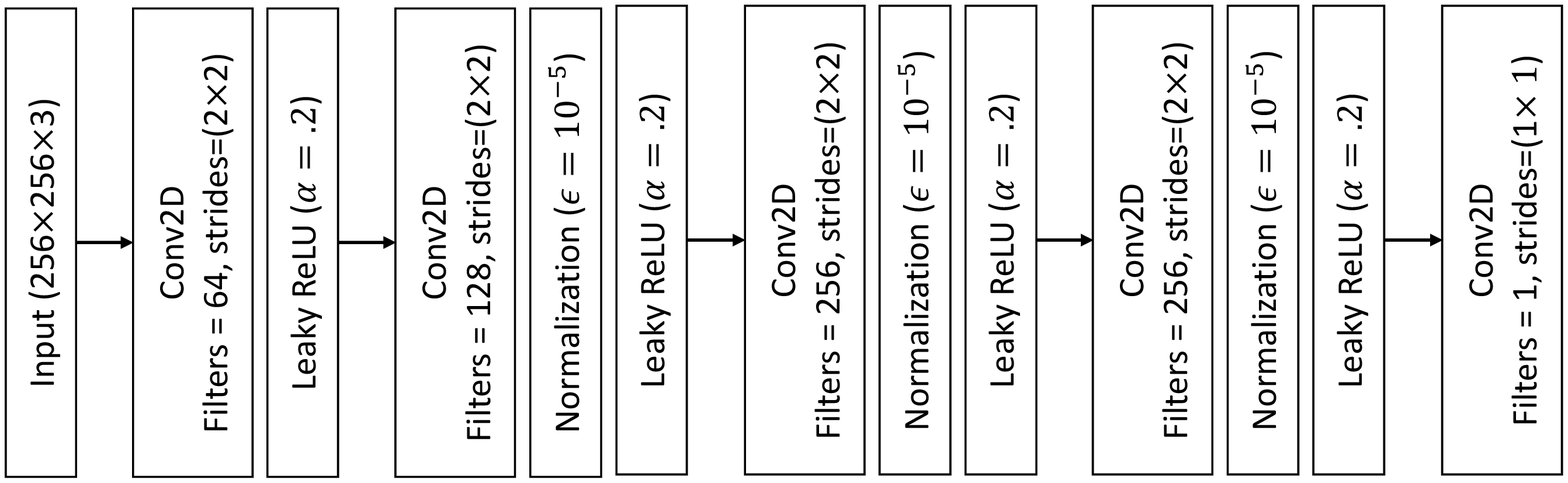}
    \caption{Discriminator model}
    \label{fig:discriminatormodel}
\end{figure}

Therefore, we use patches from the image and discriminate each patch. Then we take an average of those values. Using this method, we can improve the training of the discriminator. The discriminator model is explained in detail in \fref{fig:discriminatormodel}. The loss for each discriminator is given in \eqref{eq:total_x_loss} and \eqref{eq:total_y_loss}.

\begin{align}
    \label{eq:total_x_loss}
    \mathcal{L}_{D_{high}} = H(D_{high}(G(S_{low})), \mathbf{0}) + H(D_{high}(S_{high}), \mathbf{1})\\
    \label{eq:total_y_loss}
    \mathcal{L}_{D_{low}} = H(D_{low}(F(S_{high})), \mathbf{0}) + H(D_{low}(S_{low}), \mathbf{1})
\end{align} 
\subsection{Summary}
The first NN alone can produce $R$,$I$ decomposition for both low light and well-lit images after training. The enhancement network performs the enhancement of the $I$ and puts together $R$ and $I$ to generate enhanced image. The second network is optimized to minimize a wide range of errors.

\section{Results and discussion}

\subsection{Performance metrics}

In general, the preferred evaluation measure ranking based on visual feedback. For this purpose, this section contains examples of the results generated by the proposed pipeline. The results were also analysed for the pixel-wise mean squared error whenever possible to provide a numerical performance evaluation. Also, we are using the Structural Similarity (SSIM) and Naturalness Image Quality Evaluator (NIQE) as the standard metrics to compare the results. Note that we are using the ratio between the NIQE values of the predicted image and the ground truth image. Furthermore, in order to validate the reasoning behind the specific components of the proposed pipeline, several results were generated to ensure how the inner workings of the system comply with our design goals (instead of treating NN as a black box). 

\subsection{Retinex decomposition}
\begin{figure}[h]
    \centering
    \includegraphics[width=0.8\linewidth]{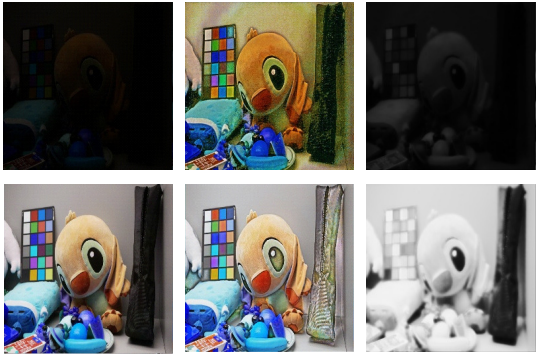}
    \caption{Results of decomposition network. Row 1 (L to R): $S_{low}, R_{low}, I_{low}$. Row 2 (L to R): $S_{high}, R_{high}, I_{high}$}
    \label{fig:decomp-result}
\end{figure}

\fref{fig:decomp-result} shows the performance of the decomposition network for a chosen image. The similarity of $R_{high} \approx R_{low}$ even when $S_{high}$ and $S_{low}$ are different is indicative of the performance of the decomposition network.

\subsection{GAN based illumination enhancement}
The GAN is used to enhance the images and illumination maps. The notation for these processes are: $\rightarrow$ denotes NNs trained under adversarial loss and $\rightleftharpoons$ denotes NNs trained under cycle consistency loss (with or without additional loss functions). Two separate experiments were done as per the following approaches. 

\begin{figure}[b]
    \centering
    \includegraphics[width=0.8\linewidth]{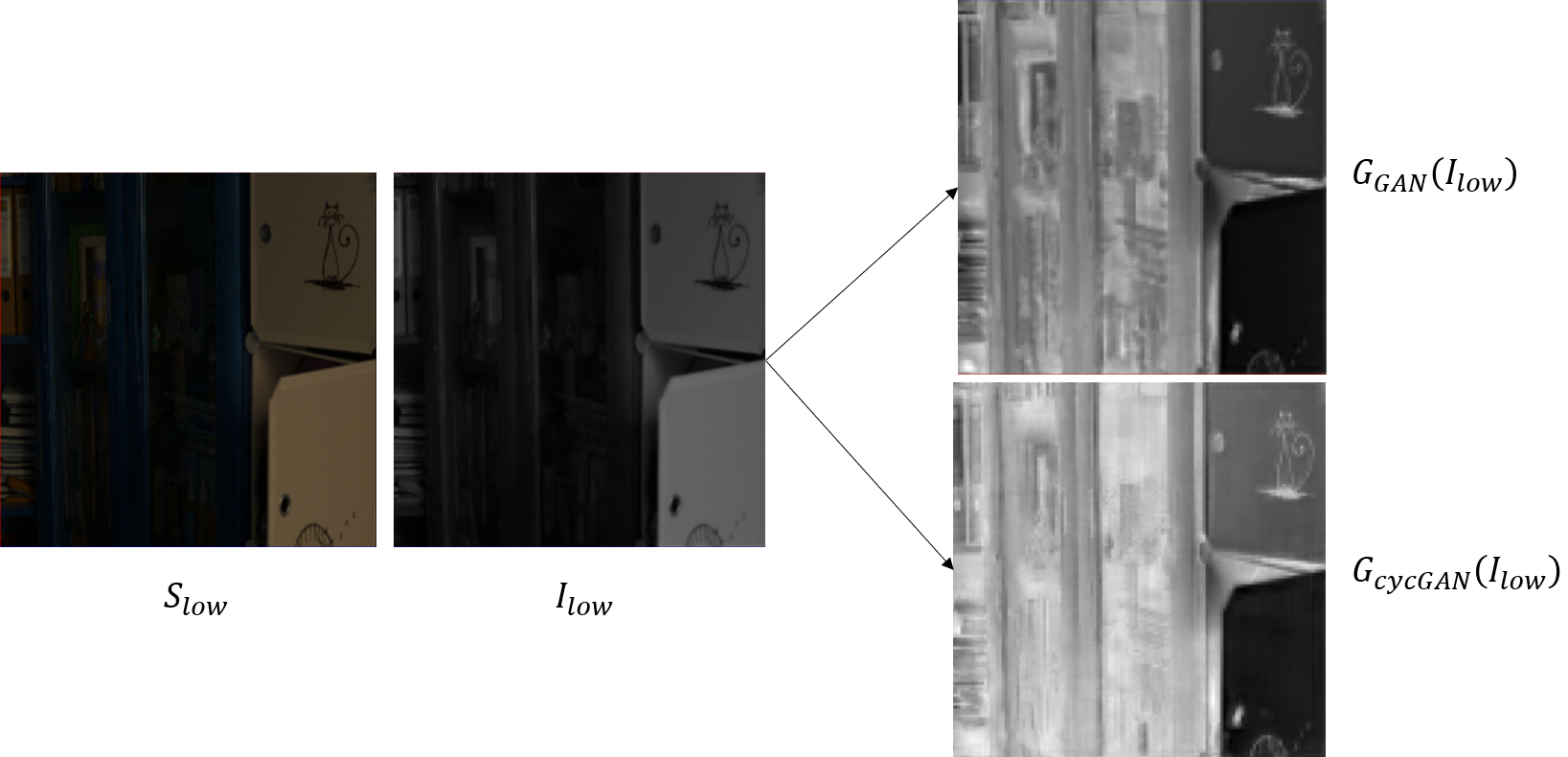}
    \caption{Comparison of the results from CycleGAN based illumination enhancement and GAN based illumination enhancement.}
    \label{fig:result-cyclegan-illumination}
\end{figure}

By training the GAN on $[R,I]$, The performance of this network is shown in \fref{fig:result-cyclegan-illumination}. It is clearly visible how CycleGAN produced a better illumination enhancement compared to generic GAN.

\subsection{Complete low-light image enhancement system}

Our work studies the results from four approaches to low light enhancement ($G$ and $G_1$ are explained in the proposed work section. $G^a$,$G^b$ are different NNs whose architecture has been kept as close as to $G$ for benchmarking).

\begin{enumerate}
    \item $G^a :S_{low} \rightarrow S_{high}$\\
    Enhancing low light images using a generic GAN.
    \item $G^b :S_{low} \rightarrow S_{high}$ with $G_1 :(R_{low},I_{Low}) \rightarrow I_{high}$ using supervised training data. Enhancing low light images using Retinexnet.
    \item $G^c :S_{low} \rightarrow S_{high}$ with $G_1 :(R_{low},I_{Low}) \rightarrow I_{high}$ as an intermediate step. Enhancing low light images using a retinex aware GAN pipeline.
    \item $G :S_{low} \rightleftharpoons S_{high}$ with $G_1 :(R_{low},I_{Low}) \rightleftharpoons I_{high}$ as an intermediate step. Enhancing low light images using a retinex aware CycleGAN pipeline.
\end{enumerate}

\begin{figure}[ht]
    \centering
    \includegraphics[width=\linewidth]{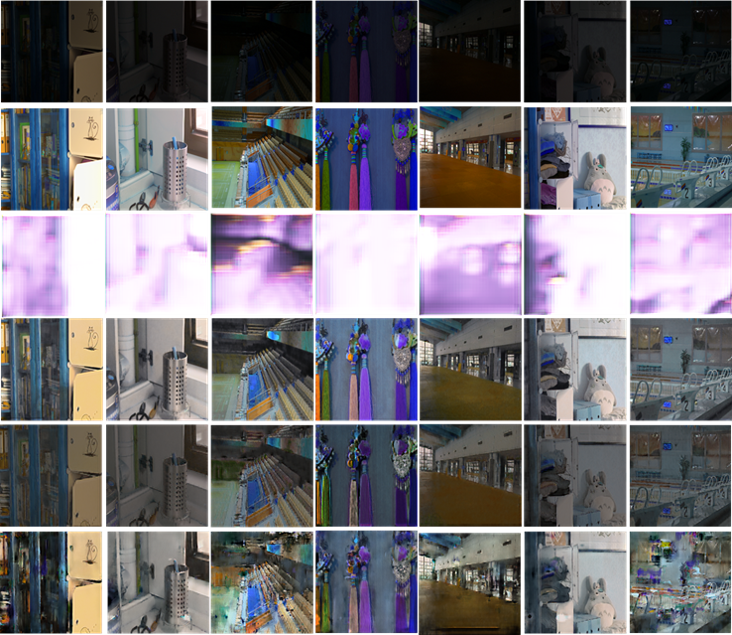}
    \caption{Results from the complete pipeline. Row 1: Low light input images. Row 2: Well-lit ground truth. Row 3: $G^a$ Generic GAN enhanced image. Row 4: $G^b$ Retinexnet enhanced image. Row 5: $G^c$ retinex based GAN enhanced image. Row 5: $G$ Retinex based CycleGAN (ours) enhanced image. 
    }
    \label{fig:results-full-pipeline}
\end{figure}

The results of all the four cases are given in \fref{fig:results-full-pipeline}. The comparison between them are given in Table. \ref{tab:numerical_results}. As shown in the table both Retinexnet and proposed method have low MSE but when compared with NIQE our method shows better results overall.

\begin{table}[t]
    \centering
    \caption{Numerical comparision of algorithm performance (MSE - lower is better, NIQE ratio - higher is better)}
    \begin{tabular}{|c|c|c|c|c|}
        \hline
         Algorithm & MSE  & NIQE ratio \\
         \hline
         Ground truth & 0.0000  & 1.0000\\
         DCGAN & 0.2171 & 1.4592\\
         Retinex based DCGAN & 0.0514 & 1.6967\\
         Retinexnet & 0.0090  & 1.7896\\
         Proposed algorithm & 0.0173  & 1.7921\\
         \hline
    \end{tabular}
    \label{tab:numerical_results}
\end{table}

\subsection{The ablation study of the inner-workings of the model.}
\begin{figure}[htb!]
    \centering
    \begin{subfigure}[t]{\linewidth}
        \centering
    	\includegraphics[width=.97\linewidth]{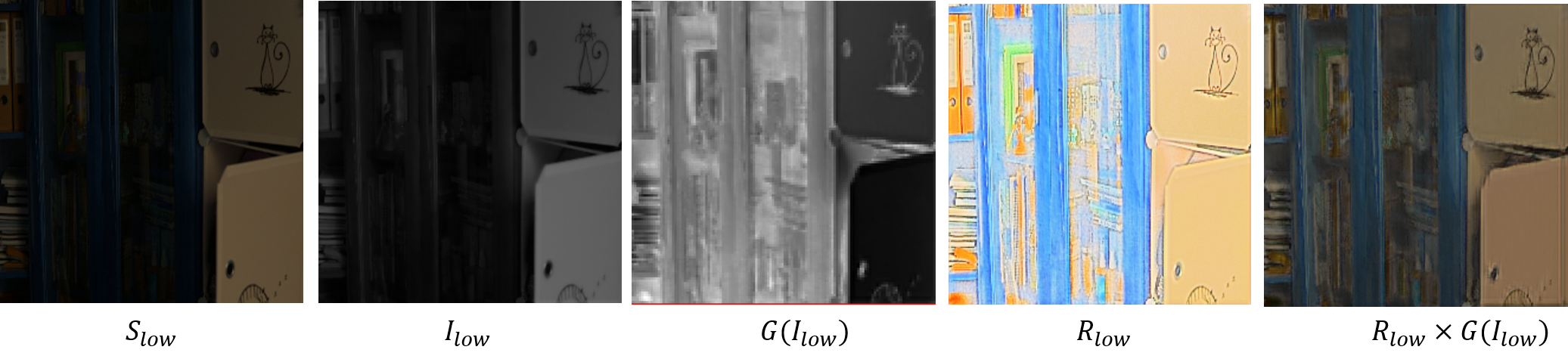}
        \caption{Layer output from GAN with retinex model.}
		\label{fig:gan_w_retinex_result}
    \end{subfigure}
    \begin{subfigure}[t]{\linewidth}
        \centering
    	\includegraphics[width=.97\linewidth]{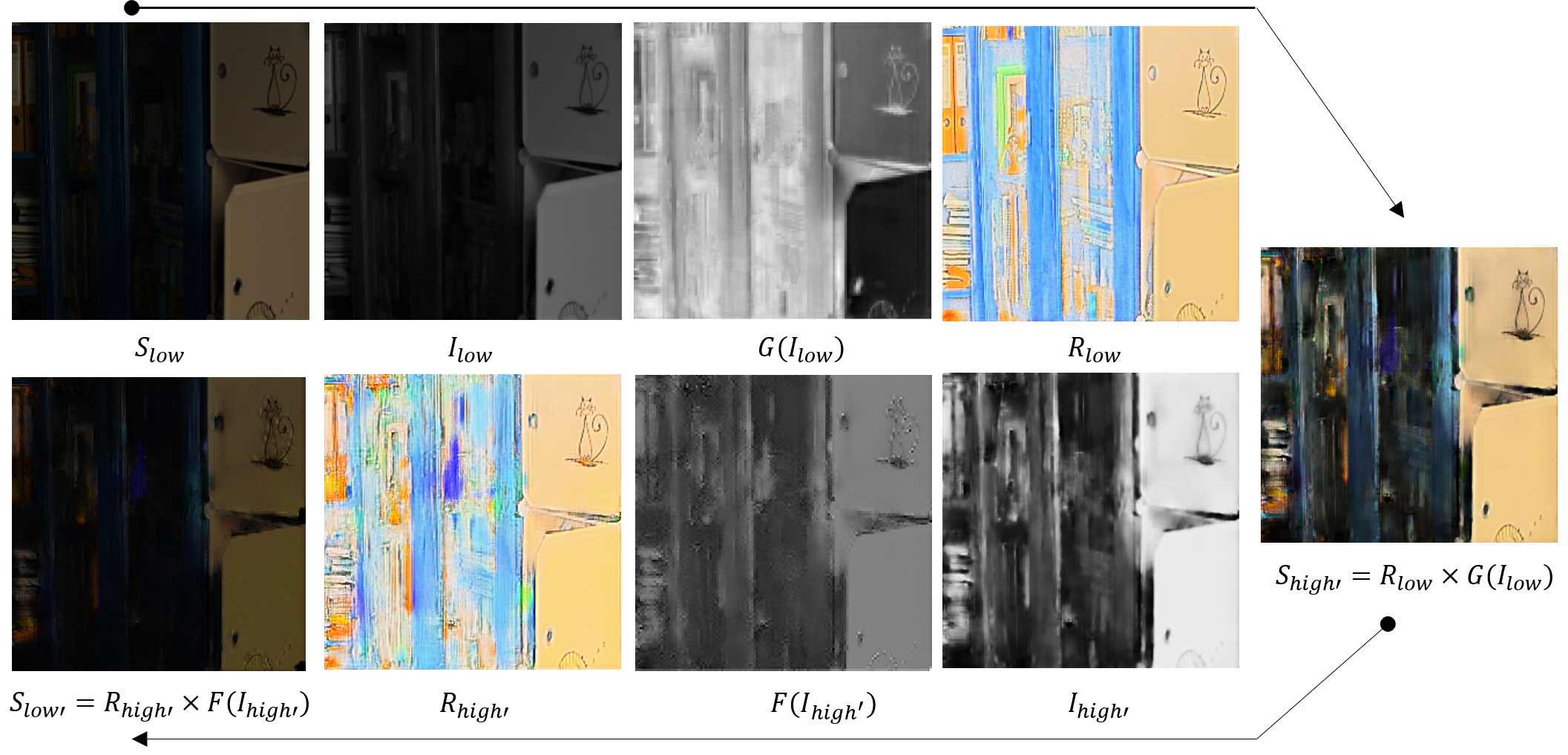}
        \caption{Layer output from Cycle-GAN with retinex model.}
		\label{fig:cyclegan_w_retinex_result}
    \end{subfigure}
    \caption{Comparison of intermediate output images.}
    \label{fig:my_label}
\end{figure}

The importance of each component of the model is analysed next. \fref{fig:gan_w_retinex_result} and \fref{fig:cyclegan_w_retinex_result} show how the NN works in the hidden layers. Layers (from left to right denoting layers from the input to the output) show how the light enhancement process happens sequentially. 
In \fref{fig:cyclegan_w_retinex_result} we observe that this results has been further enhanced through cycle consistency loss.
In \fref{fig:cyclegan_w_retinex_result}, the similarity of the reflectance of the low light image ($R_{low}$) and well-lit image ($R_{high}'$) highlights the functionality of the cycle consistency loss we introduced. We observe further enhancement in the output $S_{high}'$ which shows that the generative model plays a significant role in image enhancement.

\section{Conclusion}
Ability to work under low light conditions is an important goal for computer vision. Among other things, enhancing the lighting in images is a crucial milestone in this. Classical image processing algorithms and traditional deep learning algorithms have their strengths and weaknesses.

This paper proposes a pipeline to overcome several weaknesses of traditional deep learning algorithms by being able to use paired and unpaired datasets and use both traditional CNN and GAN architectures in a collaborative manner.

\section{Future work}
Even though this paper proposes a methodology to use both paired and unpaired datasets in the enhancement pipeline, the individual components use only one type of dataset. The future work should try to use both types of datasets in each step of the pipeline. The enhanced images of this work shows issues with respect to smoothness. Future work should explore the possibility of enhancing images while preserving natural-like smoothness.

The light enhancement problem is studied in this work as a two-class problem -- low light and well-lit. The future work should try to realize different discrete levels of lighting enhancement followed by continuous levels of lighting enhancement. Furthermore, object detection, segmentation, captioning, etc. for low light images could be built upon this work.

\section*{Acknowledgment}
This research was made possible through the contribution from the citizens of Sri Lanka towards the state-funded university system.

\bibliographystyle{IEEEtran}
\bibliography{references}

\begin{thebibliography}{10}
\providecommand{\url}[1]{#1}
\csname url@samestyle\endcsname
\providecommand{\newblock}{\relax}
\providecommand{\bibinfo}[2]{#2}
\providecommand{\BIBentrySTDinterwordspacing}{\spaceskip=0pt\relax}
\providecommand{\BIBentryALTinterwordstretchfactor}{4}
\providecommand{\BIBentryALTinterwordspacing}{\spaceskip=\fontdimen2\font plus
\BIBentryALTinterwordstretchfactor\fontdimen3\font minus
  \fontdimen4\font\relax}
\providecommand{\BIBforeignlanguage}[2]{{%
\expandafter\ifx\csname l@#1\endcsname\relax
\typeout{** WARNING: IEEEtran.bst: No hyphenation pattern has been}%
\typeout{** loaded for the language `#1'. Using the pattern for}%
\typeout{** the default language instead.}%
\else
\language=\csname l@#1\endcsname
\fi
#2}}
\providecommand{\BIBdecl}{\relax}
\BIBdecl

\bibitem{computer-vision-survey}
X.~Feng, Y.~Jiang, X.~Yang, M.~Du, and X.~Li, ``Computer vision algorithms and
  hardware implementations: A survey,'' \emph{Integration}, vol.~69, pp.
  309--320, 2019.

\bibitem{Land1977}
E.~H. Land, ``{The Retinex Theory of Color Vision},'' \emph{Sci. Am.}, vol.
  237, no.~6, pp. 108--128, 1977.

\bibitem{recent-advances-cnn}
J.~Gu, Z.~Wang, J.~Kuen, L.~Ma, A.~Shahroudy, B.~Shuai, T.~Liu, X.~Wang,
  G.~Wang, J.~Cai \emph{et~al.}, ``Recent advances in convolutional neural
  networks,'' \emph{Pattern Recognition}, vol.~77, pp. 354--377, 2018.

\bibitem{gan}
I.~Goodfellow, J.~Pouget-Abadie, M.~Mirza, B.~Xu, D.~Warde-Farley, S.~Ozair,
  A.~Courville, and Y.~Bengio, ``Generative adversarial nets,'' in
  \emph{Advances in neural information processing systems}, 2014, pp.
  2672--2680.

\bibitem{AHE}
S.~M. Pizer, E.~P. Amburn, J.~D. Austin, R.~Cromartie, A.~Geselowitz, T.~Greer,
  B.~ter Haar~Romeny, J.~B. Zimmerman, and K.~Zuiderveld, ``Adaptive histogram
  equalization and its variations,'' \emph{Computer vision, graphics, and image
  processing}, vol.~39, no.~3, pp. 355--368, 1987.

\bibitem{grad_enhance}
M.~Tanaka, T.~Shibata, and M.~Okutomi, ``Gradient-based low-light image
  enhancement,'' in \emph{2019 IEEE International Conference on Consumer
  Electronics (ICCE)}.\hskip 1em plus 0.5em minus 0.4em\relax IEEE, 2019, pp.
  1--2.

\bibitem{land1983recent}
E.~H. Land, ``Recent advances in retinex theory and some implications for
  cortical computations: color vision and the natural image.''
  \emph{Proceedings of the National Academy of Sciences of the United States of
  America}, vol.~80, no.~16, p. 5163, 1983.

\bibitem{wang2015variational}
W.~Wang and C.~He, ``A variational model with barrier functionals for
  retinex,'' \emph{Siam Journal on Imaging Sciences}, vol.~8, no.~3, pp.
  1955--1980, 2015.

\bibitem{ret_noise}
M.~Li, J.~Liu, W.~Yang, X.~Sun, and Z.~Guo, ``Structure-revealing low-light
  image enhancement via robust retinex model,'' \emph{IEEE Transactions on
  Image Processing}, vol.~27, no.~6, pp. 2828--2841, 2018.

\bibitem{lime}
X.~Guo, Y.~Li, and H.~Ling, ``{LIME}: Low-light image enhancement via
  illumination map estimation,'' \emph{IEEE Transactions on image processing},
  vol.~26, no.~2, pp. 982--993, 2016.

\bibitem{lecun1999object}
Y.~LeCun, P.~Haffner, L.~Bottou, and Y.~Bengio, ``Object recognition with
  gradient-based learning,'' in \emph{Shape, contour and grouping in computer
  vision}.\hskip 1em plus 0.5em minus 0.4em\relax Springer, 1999, pp. 319--345.

\bibitem{Ballard1987ModularLI}
D.~H. Ballard, ``Modular learning in neural networks,'' in \emph{AAAI}, 1987.

\bibitem{imagenet}
A.~Krizhevsky, I.~Sutskever, and G.~E. Hinton, ``Imagenet classification with
  deep convolutional neural networks,'' in \emph{Advances in neural information
  processing systems}, 2012, pp. 1097--1105.

\bibitem{cnn1}
L.~Xu, J.~Ren, C.~Liu, and J.~Jia, ``Deep convolutional neural network for
  image deconvolution,'' \emph{Advances in Neural Information Processing
  Systems}, vol.~2, pp. 1790--1798, 01 2014.

\bibitem{llnet}
K.~G. Lore, A.~Akintayo, and S.~Sarkar, ``{LLNet}: A deep autoencoder approach
  to natural low-light image enhancement,'' \emph{Pattern Recognition},
  vol.~61, pp. 650--662, 2017.

\bibitem{llcnn}
L.~Tao, C.~Zhu, G.~Xiang, Y.~Li, H.~Jia, and X.~Xie, ``{LLCNN}: A convolutional
  neural network for low-light image enhancement,'' in \emph{2017 IEEE Visual
  Communications and Image Processing (VCIP)}.\hskip 1em plus 0.5em minus
  0.4em\relax IEEE, 2017, pp. 1--4.

\bibitem{msrnet}
L.~Shen, Z.~Yue, F.~Feng, Q.~Chen, S.~Liu, and J.~Ma, ``Msr-net: Low-light
  image enhancement using deep convolutional network,'' \emph{arXiv preprint
  arXiv:1711.02488}, 2017.

\bibitem{pipeline-nn}
Y.~Guo, X.~Ke, J.~Ma, and J.~Zhang, ``A pipeline neural network for low-light
  image enhancement,'' \emph{IEEE Access}, vol.~7, pp. 13\,737--13\,744, 2019.

\bibitem{deep-retinex-decomp}
C.~Wei, W.~Wang, W.~Yang, and J.~Liu, ``Deep retinex decomposition for
  low-light enhancement,'' in \emph{BMVC}, 2018.

\bibitem{see_in_dark}
C.~Chen, Q.~Chen, J.~Xu, and V.~Koltun, ``Learning to see in the dark,'' in
  \emph{Proceedings of the IEEE Conference on Computer Vision and Pattern
  Recognition}, 2018, pp. 3291--3300.

\bibitem{cycle-gan}
J.-Y. Zhu, T.~Park, P.~Isola, and A.~A. Efros, ``{Unpaired image-to-image
  translation using cycle-consistent adversarial networks},'' in
  \emph{Proceedings of the IEEE international conference on computer vision},
  2017, pp. 2223--2232.

\bibitem{en-gan}
\BIBentryALTinterwordspacing
Y.~Jiang, X.~Gong, D.~Liu, Y.~Cheng, C.~Fang, X.~Shen, J.~Yang, P.~Zhou, and
  Z.~Wang, ``{EnlightenGAN: Deep Light Enhancement without Paired
  Supervision},'' \emph{Preprint}, 2019. [Online]. Available:
  \url{http://arxiv.org/abs/1906.06972}
\BIBentrySTDinterwordspacing

\bibitem{brightening-train}
W.~Wang, C.~Wei, W.~Yang, and J.~Liu, ``{GLADNet: Low-Light Enhancement Network
  with Global Awareness},'' \emph{2018 13th IEEE International Conference on
  Automatic Face \& Gesture Recognition (FG 2018)}, pp. 751--755, 2018.

\bibitem{attention_guide}
F.~Lv and F.~Lu, ``Attention-guided low-light image enhancement,'' \emph{arXiv
  preprint arXiv:1908.00682}, 2019.

\bibitem{Exdark}
Y.~P. Loh and C.~S. Chan, ``Getting to know low-light images with the
  exclusively dark dataset,'' \emph{Computer Vision and Image Understanding},
  vol. 178, pp. 30--42, 2019.

\bibitem{adam}
D.~P. Kingma and J.~Ba, ``{Adam: A method for stochastic optimization},''
  \emph{arXiv preprint arXiv:1412.6980}, 2014.

\bibitem{unet}
O.~Ronneberger, P.~Fischer, and T.~Brox, ``{U-Net: Convolutional Networks for
  Biomedical Image Segmentation},'' in \emph{Medical Image Computing and
  Computer-Assisted Intervention -- MICCAI 2015}, N.~Navab, J.~Hornegger, W.~M.
  Wells, and A.~F. Frangi, Eds., vol. 9351.\hskip 1em plus 0.5em minus
  0.4em\relax Springer International Publishing, 2015, pp. 234--241.

\end{thebibliography}

\end{document}